\documentclass[a4paper,twocolumn]{Gaia2004} 
\usepackage{times}      
\usepackage{epsfig}     
\usepackage{natbib}     
\title{Theoretical modelling of late-type giant atmospheres: preparing for Gaia}

\author[1,2]{A.~Ku\v{c}inskas}
\affil[1]{National Astronomical Observatory of Japan, 2-21-1 Mitaka, Tokyo, 181-8588, Japan}
\affil[2]{Institute of Theoretical Physics and Astronomy, Go\v{s}tauto 12, Vilnius 01108, Lithuania}

\author[3,4]{I.~Brott}
\author[3]{P.H.~Hauschildt}
\affil[3]{Hamburger Sternwarte, Gojenbergsweg 112, 21029 Hamburg, Germany}
\affil[4]{INTEGRAL Science Data Centre, Chemin d'Ecogia 16, 1290 Versoix, Switzerland}

\author[5]{H.G.~Ludwig}
\author[5]{L.~Lindegren}
\affil[5]{Lund Observatory, Lund University, Box 43, SE-221 00, Lund, Sweden}

\author[6]{T.~Tanab\'{e}}
\affil[6]{Institute of Astronomy, The University of Tokyo, Mitaka,
Tokyo 181-0015, Japan}

\author[7]{V.~Vansevi\v{c}ius}
\affil[7]{Institute of Physics, Savanoriu 231, Vilnius 02300,
Lithuania}

\bibpunct{(}{)}{;}{a}{}{,}  

\begin{document}


\maketitle

\begin{abstract}

Late type giants (RGB/AGB stars) will be important tracers of the
Galactic morphology and evolution in the framework of Gaia, as
they are intrinsically bright and thus can probe distant stellar
populations or those obscured by interstellar extinction.  A
realistic representation of their atmospheres and spectra with
stellar atmosphere models is thus of crucial importance, both for
the design and optimization of Gaia instruments, as well as the
interpretation of provided astrophysical data. Our analysis of
synthetic photometric colors of late-type giants based on {\tt
PHOENIX}, {\tt MARCS} and {\tt ATLAS} model atmospheres indicates
a general agreement between the current theoretical predictions
and observations in the framework of stationary 1-D model
atmospheres. Presently available models allow temperature
determinations of RGB/AGB stars to an accuracy of $\sim\pm100\,K$.
In an exploratory study we try to quantify possible residual
systematic effects due to the approximations made in 1-D models
using full 3-D hydrodynamical models. We find that differences in
broad-band photometric colors calculated with 1-D and 3-D models
are significant, translating to the offsets in effective
temperature of up to $\Delta T_{\rm eff}\sim70$\,K. Clearly, full
3-D hydrodynamical models will help to alleviate such ambiguities
in current theoretical modeling. Additionally, they will allow to
study new phenomena, to open qualitatively new windows for stellar
astrophysics in the Gaia-era.

\end{abstract}

\section{Importance of late-type giants in the context of Gaia}

Gaia will be an excellent instrument for studying Galactic stellar
populations, especially in terms of their detailed formation
histories, as it will provide parallaxes, proper motions, radial
velocities, and astrophysical parameters (effective temperatures,
metalicities, gravities and reddenings) for stars down to Gaia
magnitude $G=20$, which translates to Johnson $V\sim21$ for
$V-I=2.0$ (or $T_{\rm eff}\sim3700$\,K) and $V\sim20.5$ for
$V-I=1.0$ ($T_{\rm eff}\sim4800$\,K). This unique combination of
kinematical and astrophysical information which is not accessible
otherwise will provide an opportunity to investigate formation and
evolution of the Galaxy in unprecedented detail.

Currently stellar populations are predominantly dated using the
main sequence turn-off point (MSTO) stars which is the most
reliable method among those available today. This requires precise
photometry of typically faint main sequence stars, i.e. those
located at least one magnitude below the MSTO (at 15 Gyr and solar
metalicity the MSTO is at $M_V\sim+4.6$). Moreover, photometric
data should be obtained for a statistically sufficient number of
stars in a given target population. Surprisingly, these two
requirements appear to be rather demanding in the context of Gaia,
since already a mild extinction in the galactic disk of e.g.
$A_V\sim0.7$\,mag/kpc will make MSTO stars too faint to be
detected with Gaia at rather short distances. The limiting
distance to which stellar population can be dated with Gaia using
MSTO stars will vary depending on the number of MSTO stars,
interstellar extinction, and other factors, but it will be
generally confined to $2.0-2.5$\,kpc \citep{B02}. Clearly, unless
other suitable and reliable tracers can be identified, the
capability of Gaia for dating stellar populations will be
restricted to a few kiloparsec within the solar neighborhood.
Late-type giants on the RGB and early-AGB can be very useful in
overcoming this limitation.  Being significantly brighter than
MSTO stars they can be easily observed to large distances (at
solar metalicity the absolute magnitude of a typical RGB star
located $\sim1.5$\,mag below the RGB tip is $M_V\sim+0.5$). Hence,
late-type giants will be extensively employed in the Gaia data
analysis as they can provide information about distant stellar
populations not accessible by using fainter main sequence stars.

Until now, however, late-type giants have been scarcely used for
dating stellar populations. Firstly, the observational scatter in
their photometric colors is typically large compared to the spread
of isochrones corresponding to different ages on the RGB/AGB, thus
uncertainties in derived ages are large. Secondly, the evolution
on the RGB/AGB is still rather uncertain, as theoretical models
are dependent on a number of poorly constrained modeling
parameters such as the mixing-length parameter and mass loss rate.

However, for Gaia there is evidence that effective temperature
determinations of late-type giants may be feasible at an accuracy
level of $1-2\%$, mostly from medium-band photometry (see Jordi 
\& H{\o}g, this volume). The quoted error level corresponds to the
formal error resulting from the fitting procedure with theoretical
model atmospheres. At least for some stars the medium-band
photometry will be supplemented by ground-based data obtained by
massive future spectroscopic (e.g.  RAVE) and photometric (e.g.
VST, VISTA, UKIDSS) surveys.  Near-infrared photometry will be
especially important in this context, as it will allow to put even
tighter constrains on the effective temperatures \citep[see][
K03]{K03} and interstellar extinction. This will allow to obtain
sufficiently precise astrophysical parameters of RGB/AGB stars
($\sigma(T_{\rm eff})\sim50-100$\,K -- see Sect.~2), which, in
turn, may provide age estimates at an (formal) accuracy level of
$\sim20\%$ or better.

The most essential ingredient missing here today is the
availability of reliable evolutionary models, which would provide
a solid basis for precise and unambiguous dating of stellar
populations with RGB/AGB stars. At least partly this is due to the
inadequate treatment of multi-dimensional phenomena (especially
convection) within the classical 1-D models. It is conceivable
that at least some of these problems will be solved with full 3--D
hydrodynamical models, which may be available for routine use
within $\sim$10 years from now. Being constructed to account for
time-dependent and non-spherical phenomena, 3-D hydrodynamical
models may help to alleviate the majority of ambiguities in
theoretical modeling of stellar evolution on the RGB/AGB
(especially those related with convection, mass-loss, pulsations,
etc.).

\section{Current status in theoretical modeling of late-type
giant atmospheres: synthetic broad-band photometric colors}

\begin{figure*}[]
\begin{center}
    \leavevmode
    \centerline{\epsfig{file=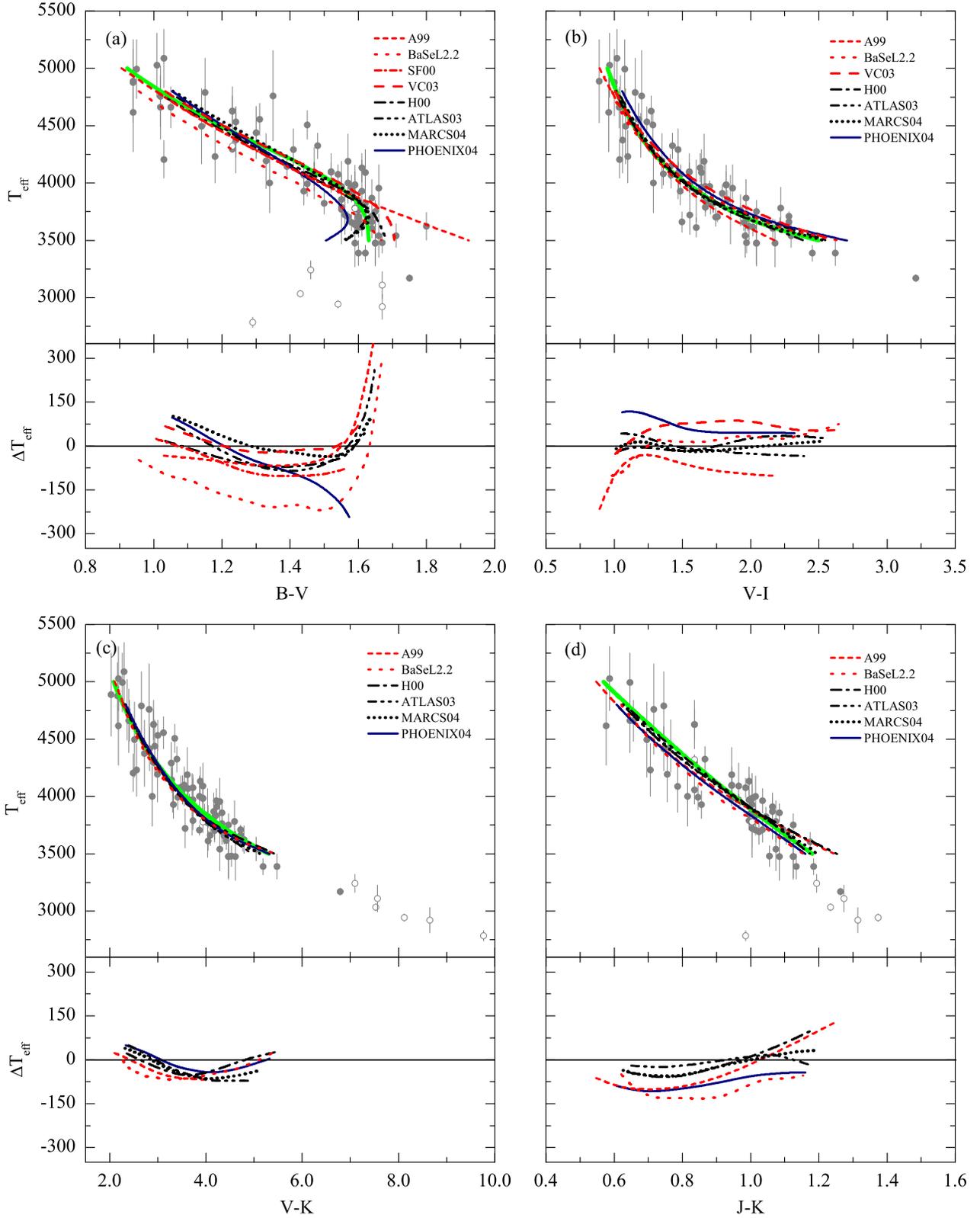,width=1.0\linewidth}}
\end{center}
\caption{Empirical and theoretical $T_{\rm eff}-{\rm color}$
relations for late-type giants in different $T_{\rm eff}-{\rm
color}$ planes (a-d, top panels). Filled circles are non-variable
late-type giants, several semiregular variables are highlighted as
open circles. Thick solid line is a best-fit to the observed
colors in a given $T_{\rm eff}-{\rm color}$ plane. Several
existing $T_{\rm eff}-{\rm color}$ relations are shown as well,
together with $T_{\rm eff}-{\rm color}$ scales constructed using
synthetic colors of {\tt PHOENIX}, {\tt MARCS} and {\tt ATLAS}
(see text for details). Bottom panels in each figure show the
difference between various $T_{\rm eff}-{\rm color}$ relations and
the best fit to the observed data in a given $T_{\rm eff}-{\rm
color}$ plane ($\Delta T_{\rm eff}=T_{\rm eff}^{\rm other}-T_{\rm
eff}^{\rm bestfit}$).} \label{}
\end{figure*}

Clearly, a number of issues related to stellar evolution on the
RGB/AGB have to be clarified before late-type giants can be widely
used for reliable dating of stellar populations. Can today's
theoretical model atmospheres provide a sufficient backup to
obtain precise astrophysical parameters of individual giants from
observations, in particular broad-band photometry?

To address this question we have performed an extensive comparison
of synthetic broad-band photometric colors of late-type giants
with observations, covering a wide range of effective
temperatures, gravities and metalicities ($T_{\rm
eff}=3000-4500$\,K, ${\rm log}\,g=0.5-2.5$ and
$[M/H]=0,-1.0,-2.0$). Synthetic spectra used for this purpose were
produced with the most recent versions of {\tt PHOENIX}, {\tt
MARCS} and {\tt ATLAS} model atmospheres (for details see: {\tt
PHOENIX}: Brott \& Hauschildt, this volume; {\tt MARCS}: Plez
2003; {\tt ATLAS}: Castelli \& Kurucz, 2003). Synthetic colors
were calculated in the Johnson-Cousins-Glass system, using filter
definitions from \citet{B90} for the Johnson-Cousins \textit{BVI}
bands and from \citet{BB88} for Johnson-Glass \textit{JK} bands. A
carefully selected sample of local late-type giants with $T_{\rm
eff}<5000$\,K was used for this comparison, with precise
determinations of effective temperature available from
interferometry (errors in $T_{\rm eff}$ typically below $5\%$). We
also made sure that stars in this sample are non-peculiar
(chemically or otherwise) and non-variable. Comparisons were made
in $T_{\rm eff}-{\rm color}$ and color$-$color planes, with
synthetic photometric colors plotted according to a $T_{\rm
eff}-{\rm log}\,g$ relation of \citet{H00}.  We included several
commonly used $T_{\rm eff}-{\rm color}$ relations in this analysis
too (Fig.1; A99: Alonso et al. 1999; BaSel2.2: Lejeune et al.
1998; SF00: Sekiguchi \& Fukugita 2000; VC03: Vandenberg \& Clem
2003; H00: Houdashelt et al. 2000). Here we present some results
for solar metalicity; a full discussion covering also sub-solar
metalicities will be given elsewhere (Ku\v{c}inskas et al. 2005,
A\&A, in preparation).

Figure~1 shows that there is a general agreement between the
synthetic broad-band colors predicted by today's stationary 1-D
model atmospheres and the observations, especially in the $T_{\rm
eff}-(V-K)$ and $T_{\rm eff}-(J-K)$ planes. Moreover, there is a
remarkable agreement among the synthetic colors calculated with
different stellar atmosphere models, translating typically into an
uncertainty in $T_{\rm eff}$ of $\pm50$\,K over a large range of
effective temperatures, gravities and metalicities. This indicates
that presently available models may allow to obtain effective
temperatures of late-type giants with a precision of $\sim \pm
100$\,K, using a single broad-band photometric index like $V-I$,
or $V-K$. The accuracy can still be improved if optical and
near-infrared broad-band fluxes (such as Johnson's
\textit{BVRIJHK}) are fitted simultaneously with synthetic fluxes
to obtain an estimate of $T_{\rm eff}$ (K03). Typically, observed
fluxes at 5-6 wavelength points are sufficient to achieve a formal
fitting accuracy (standard deviation) of $\sigma(T_{\rm
eff})\sim50$\,K.  This considerably reduces the observational
scatter on the RGB/AGB, allowing to obtain ages with a precision
of at least $\pm20\%$ (see K03 for an example with LMC/SMS
clusters).

\section{Towards higher accuracy: full 3-D hydrodynamical model atmospheres}

\begin{figure}[]
\begin{center}
    \leavevmode
    \centerline{\epsfig{file=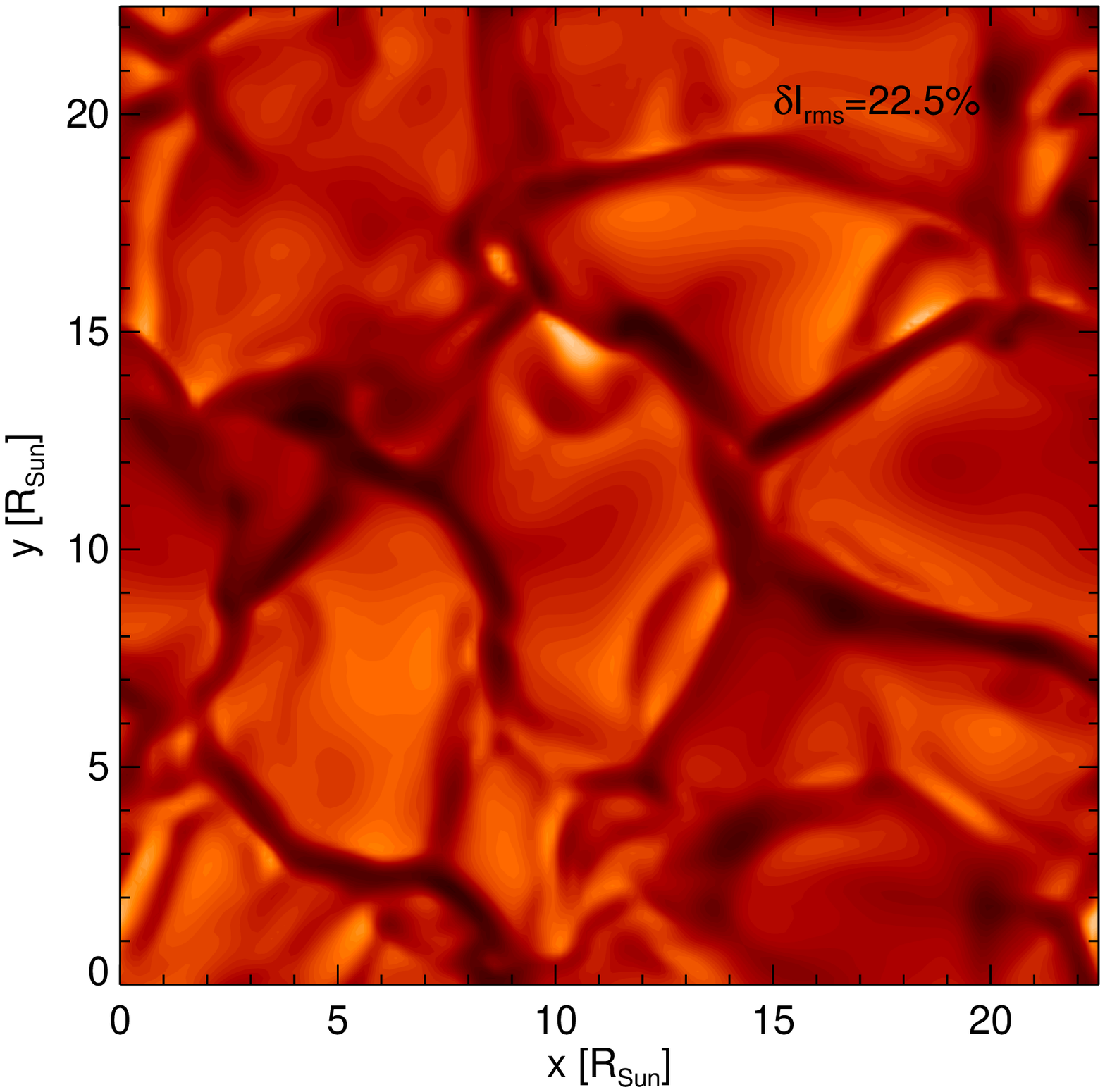,width=0.8\linewidth}}
\end{center}

\vspace{-\baselineskip}

\caption{Snapshot of the emergent white light intensity during the
  temporal evolution of convective flows on the surface of a red
  giant ($T_{\rm eff}=3700$\,K, ${\rm log}\,g=1.0$, $[M/H]=0$). A
  typical granular pattern is discernible.}
\label{granulation}
\end{figure}

We conducted an exploratory study (Ludwig et al. 2005, in
preparation) to estimate systematic effects on broad-band colors
related to changes of the mean temperature profile and the
presence of temperature inhomogeneities in a fully time-dependent
3-D hydrodynamical model atmosphere relative to a 1-D standard
one. We considered a prototypical late-type giant with $T_{\rm
eff}=3700$\,K, ${\rm log}\,g=1.0$, $[M/H]=0$. It is an interesting
example since standard models predict that convection is
restricted to the optically thick layers, hence has no influence
on the atmospheric structure and consequently on the spectroscopic
or photometric stellar properties. In contrast (see
Fig.~\ref{granulation}), our hydrodynamical model predicts that
convection affects the atmospheric layers, and a typical
granulation pattern is present. As it turns out it is the result
of intense convective overshooting.

\begin{figure}[t]
\begin{center}
    \leavevmode
    \centerline{\epsfig{file=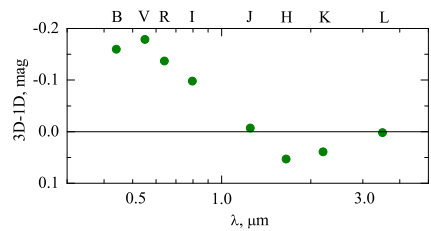,width=0.7\linewidth}}
\end{center}

\vspace{-\baselineskip}

\caption{Influence of surface granulation on the broad-band
photometric colors of a red-giant, as reflected by differences between
the predictions of 3-D full hydrodynamical and classical 1-D model
atmospheres. Note that differences in color indices -- in this
representation directly given by the magnitude differences of the two
selected bands -- are significant, e.g. $\Delta (V-K)\sim0.2$ would
correspond to $\Delta T_{\rm eff}\sim70$\,K.} \label{colorbias}
\end{figure}

1-D model atmosphere used in this comparison employed the same
physical input data (opacities, equation of state, description of
radiative transfer) as the hydrodynamical model. The convective
energy transport was treated with mixing-length theory, turbulent
pressure was neglected. Both aspects are of minor importance here
since neither process influences the atmospheric structure of the
1-D model. For the 3-D model, direct calculation of broad-band
colors is computationally too demanding, because of the complex
geometry and time-dependence of the convective flow, and a large
number of wavelength points necessary for a realistic description
of the stellar spectrum. We instead approximated the
hydrodynamical flow structure by classifying horizontal locations
according to their white light emergent intensity, by sorting
darker and increasingly brighter areas into nine intensity groups.
For each group we calculated the average (in space and time)
vertical thermal structure. The resulting nine thermal structures
were treated as standard plane-parallel model atmospheres (the
so-called 1.5-D approximation) in the subsequent spectral
synthesis calculations. The resulting radiation fields were added
according to the surface area occupied by their intensity group.
Figure~\ref{colorbias} shows the magnitude differences between
broad-band fluxes of the 3-D and 1-D model. We find color biases
corresponding to temperature offsets of up to $\Delta T_{\rm
eff}\sim70$\,K.  This is a preliminary result obtained for a
particular case but it illustrates limitations in the accuracy
which we can in principle expect when interpreting observations in
the framework of hydrostatic 1-D models. By the time Gaia is
flying we expect that hydrodynamical model atmospheres are
generally available for overcoming such limitations.

3-D hydrodynamical model atmospheres will also allow to account
for a number of qualitatively new effects. We find, for instance,
that surface granulation due to convective overshoot in the upper
photospheric layers will induce a luminosity variability on a
peak-to-peak level of $\sim0.01$\,mag, on a time scales of one
week. Additionally, intensity fluctuations on the stellar surface
will also cause a stochastic movement of the stellar photocenter
on the same time scale \citep{SL05}. Almost certainly, such
phenomena will be routinely studied in 10--15 years from now, thus
providing new windows for looking inside stars.

\section*{Acknowledgments}

\vspace{-\baselineskip}

AK acknowledges Research Fellowship of the National Astronomical
Observatory of Japan. This work was also supported by the
Wenner-Gren Foundations.

\end{document}